\documentstyle[12pt]{article}
\def\st{\scriptstyle}

\def\beq{\begin{equation}}
\def\eeq{\end{equation}}
\def\bea{\begin{eqnarray}}
\def\eea{\end{eqnarray}}

\def\ket#1{\left\vert #1\right\rangle}
\def\nor#1{\left\langle #1 | #1 \right\rangle}

\def\w1{$ W_{1+\infty} $ }
\def\wi{$ W_{\infty} $ }
\def\nn{\nonumber}
\def\ket#1{\left\vert #1\right\rangle}

\def\spq{$ {\cal S}_{ps} $\ }
\begin{document}
%
\begin{center}

\vfill
{\large Isospectral Hamiltonians and $W_{1+\infty}$ algebra}

\vspace{1.5cm}

N. Aizawa$^*$ 
\vspace{0.5cm}

{\em Department of Applied Mathematics}

{\em Osaka Women's University, Sakai, Osaka 590, Japan}

\vspace{0.5cm}
and

\vspace{0.5cm}
Haru-Tada Sato$^{\dagger}$
\vspace{0.5cm}

{\em The Niels Bohr Institute, University of Copenhagen}

{\em Blegdamsvej 17, DK-2100 Copenhagen, Denmark}

\end{center}

\vfill
\begin{abstract}
We discuss a spectrum generating algebra in the supersymmetric quantum 
mechanical system which is defined as a series of solutions to a specific 
differential equation. All Hamiltonians have equally spaced eigenvalues,  
and we realize both positive and negative mode generators of 
a subalgebra of $W_{1+\infty}$ without use of negative power of 
 raising/lowering operators of the system. 
All features in the supersymmetric case are generalized to 
the parasupersymmetric systems of order 2.
\end{abstract}

\noindent-------------------------------------------------------------------\\
$^{\dagger}$ {\footnotesize Fellow of the Danish Research Academy } \\
\mbox{}\hspace{0.3cm}{\footnotesize E-mail~: sato@nbi.dk} \\
$^*$ {\footnotesize E-mail~: aizawa@appmath.osaka-wu.ac.jp}
\newpage
%
%
\section{Introduction}
\indent

We study spectrum generating algebras in para-/supersymmetric 
quantum mechanical (PSQM/SQM) systems with equally spaced energy 
eigenvalues (For a review on supersymmetric quantum mechanics, see 
\cite{sqm}). In this article, we adopt the formalism \cite{rr} 
that we can define a series of Hamiltonians as well as raising/lowering 
operators to every solution of a specific nonlinear differential equation 
\cite{bm}-\cite{ng}. Supersymmetry is an important concept nowadays, 
however it is generally difficult to explicitly construct raising/lowering 
operators for eigenvalues separated by inhomogeneous intervals. To open a 
more general and complicated energy spectrum, it may be useful to 
investigate various properties of the spectrum generating algebra of our 
systems. Our algebra, generated by Hamiltonian and raising/lowering 
operators, is not a finite dimensional one like 
the harmonic oscillator (HO).

The importance of infinite dimensional Lie algebras in theoretical 
physics has increased since the discovery of Kac-Moody and 
Virasoro algebras. Among other things, \w1 algebra and its subalgebras 
have become popular in recent years in various subjects \cite{wsym} as  
they describe symmetries of systems. As an interesting connection to 
this stream, we report that subalgebras of \w1 take part in 
a spectrum generating algebra, not in a symmetry algebra.


First, we discuss the similar algebraic structure to the HO system in the SQM 
Hamiltonian with equally spaced eigenvalues. Second, a different structure 
from HO algebra is discussed in relevance to \wi algebra. 
Finally, our scheme is generalized to the PSQM system. 

%
\setcounter{equation}{0}
\section{Definition of \w1 and \wi}
\indent

Let us here note the definitions of  \w1 and \wi algebras. 
\w1 algebra is defined as a central extension of Lie algebra of 
(higher order) differential operators on a circle. Let $ z $ be a point 
on a circle and $ D = z \st{d \over dz} $, then the commutation relations 
of the algebra generated by 
$ \{ z^n D^m\ | \  \ n, m \in {\rm\bf Z}, \ m \ge 0 \} $ 
are easily obtained. 
For the arbitrary polynomials 
of $D$, $f$ and $g$ (which may include a constant term), the commutation 
relation of \w1 is \cite{kr}
\bea
   & & [ W(z^n f(D)), \; W(z^m g(D)) ]   \nn \\ 
   & & = W(z^{n+m}f(D+m) g(D)) - W(z^{n+m}f(D) g(D+n)) 
       + c \Psi(z^nf(D), z^mg(D)),    \label{two}
\eea
where $ W(z^n D^m ) $ is a generator corresponding to $ z^n D^m $ 
and $ c $ is called a central charge.  $ \Psi $ is the {\em 2-cocycle} 
given by
\beq
 \Psi(z^n f(D), z^m g(D)) = \delta_{n+m, 0} \left\{
\begin{array}{cc}
   \displaystyle{\sum_{j=1}^n \; f(-j) g(n-j)} & {\rm for}\ \  n \ge 1 \\
   -\displaystyle{\sum_{j=1}^m \; f(m-j) g(-j)} & {\rm for}\ \  m \ge 1.
\end{array} \right.
                                                           \label{three}
\eeq

\wi algebra is a subalgebra of \w1 and is generated by 
$ W(z^n f(D) D) $, ($n \in {\rm \bf Z} $). In other words, \wi algebra 
is obtained by subtracting an infinite number of generators 
$ W(z^n) $  ($ n = \pm 1, \pm 2, \cdots $) from \w1. The unitary 
representations of these algebras have been discussed in \cite{kr} and 
\cite{afmo}.

%
\setcounter{equation}{0}
\section{Hamiltonians with equally spaced eigenvalues}
\indent

It is known that a SQM system is described by a pair of Hamiltonians 
written in the following factorized form \cite{witten}
\bea
  & & A^{\dagger} = \textstyle{1 \over \sqrt 2} ( -ip + w(x)), \quad 
      A = \textstyle{1 \over \sqrt 2} ( ip + w(x)),
                    \quad p=-i{d\over dx},            \nn \\
  & & H_- \equiv A^{\dagger} A = 
         \textstyle{1 \over 2} (p^2 + w(x)^2 - w'(x)), \label{four} \\
  & & H_+ \equiv A A^{\dagger} = 
            \textstyle{1 \over 2} (p^2 + w(x)^2 + w'(x)),  \nn
\eea
and possesses the following three important properties:  
(i) the eigenvalues of $ H_{\pm} $ are positive or zero; (ii) the ground state 
of $ H_- $ given by $ \ket G \ \propto \ \exp(-\int_{-\infty}^x w(t) dt) $ 
has zero energy because it is defined as the state annihilated by 
the operator $ A $;  
(iii) $(n+1)$-th eigenstate of $ H_- $ and $n$-th one of $ H_+ $ are 
transformed into each other by $ A $ and $A^{\dagger} $, and hence  
$H_{\pm}$ have the same energy spectra except for the ground state $\ket G$. 
All these properties stem from the factorized forms of $ H_{\pm} $. 

We suppose the following condition throughout this paper to discuss 
 Hamiltonians with equally spaced eigenvalues
\beq
        w(x)^2 + w'(x) = x^2 + k,                   \label{five}
\eeq
where $H_+$ becomes a harmonic oscillator 
and the constant $k$ yields the lowest energy of $ H_+ $
(the energy gap between $\ket{G}$ and $\ket{0}$) 
\beq
  H_+ \ket 0 = E_0 \ket 0,   \qquad     E_0 = (k+1)/2. 
\eeq
The case $ k = 1 $ is discussed in \cite{bm,fhn,rr,ng}. 
For $ k = 1 $, it is known that equation (\ref{five}) has a 
series of non-trivial solutions 
which have been found analytically(see Appendix). 
For $ k \neq 1 $, solutions could be found numerically, at least. 
We generalize their arguments independently of the value of $k$. 
Our prescription will clarify the role of $k$ as central extensions of a 
spectrum generating algebra of $H_-$. 
  
First, we mention general features of our spectrum generating algebra.
Eq. (\ref{five}) is a nonlinear differential equation with respect to 
$w(x)$. Owing to property (iii), we can consider $H_-$ as a Hamiltonian 
with equally spaced eigenvalues for each solution of (\ref{five}). 
In fact, we obtain the raising and lowering operators for excited states 
of $ H_- $
\beq
  O^{\dagger} = A^{\dagger} b^{\dagger} A, \qquad
  O = A^{\dagger} b A,                                   \label{six}
\eeq
which satisfy
\beq
 [ H_-,\; O^{\dagger} ] = O^{\dagger}, \qquad
 [ H_-, \; O ] = -O,                                   \label{seven}\\
\eeq
where $b$ and $b^{\dagger}$ are harmonic oscillator's raising/lowering 
operators in $H_+$
\beq
b^{\dagger}={1\over \sqrt{2}}(-ip +x),\quad
b ={1\over \sqrt{2}}(ip +x).
\eeq
Note that the ground state of $ H_- $ is annihilated by these
\beq
  O^{\dagger} \ket G = O \ket G = H_- \ket G = 0.           \label{eight}
\eeq
These relations are summarized in Figure 1. 

Similar to the case of usual harmonic oscillator, we can construct the 
Fock representation and coherent states (eigenstates of the lowering operator). 
The $(n+1)$-th excited state 
$\ket{\psi_n}$ of $ H_- $ starting from its first excited state is
\beq
  \ket{\psi_n} = \displaystyle{\left( \prod_{t=0}^{n-1}\; \xi_t \right)^{-1/2}}
   (O^{\dagger})^n \ket{\psi_0},                             \quad
   \xi_t = \displaystyle{ \sum_{m=0}^t\; \{ 3(E_0 + m)^2 - k(E_0+m) \} },
                                                            \label{nine} 
\eeq
which is orthonormal and satisfies 
\beq
  O^{\dagger} \ket{\psi_n} = \sqrt{\xi_n} \ket{\psi_{n+1}}, \qquad
  O\; \ket{\psi_n} = \sqrt{\xi_{n-1}} \ket{\psi_{n-1}}.
\eeq
The coherent state, which has been 
obtained for $ k=1 $\cite{fhn}, is generalized to any value of $k$;
\bea 
 & & O \ket{\alpha} = \alpha \ket{\alpha},          \nn \\
 & & \ket{\alpha} = \sum_{n=0}^{\infty} \left(
     \prod_{t=0}^{n-1} \xi_t
     \right)^{-1/2}\; \alpha^n \ket{\psi_n}.   \label{cohe}
\eea
The norm of this state converges for any value of $ |\alpha|^2 $ 
\beq
  \nor{\alpha} = \sum_{n=0}^{\infty}\; |\alpha|^{2n} \left(
   \prod_{t=0}^{n-1} \xi_t \right)^{-1}_,              \label{norm}
\eeq
regarding the power series of $ |\alpha|^2 $ where the radius of 
convergence $ R $ is infinity 
\beq
   R^{-1} = \lim_{n \rightarrow \infty} \;\left| \;
   \left( \prod_{t=0}^n \xi_t \right)^{-1}\; 
   \left( \prod_{t=0}^{n-1} \xi_t \right) \; 
   \right| 
   = 0.                                               \label{conv}
\eeq

%
\setcounter{equation}{0}
\section{Relationship to Subalgebras of \w1}
\indent

 Next, we focus our attention on the infinite dimensionality 
of the algebraic relation of (\ref{six}) and (\ref{seven}) for $H_-$. This 
property is different from the finite dimensionality of the harmonic 
oscillator $H_+$. The operators $ O^{\dagger} $ and $ O $ satisfy the 
following relations
\bea
  & & OO^{\dagger} = H_-^3 + \textstyle{1 \over 2} (3-k) H_-^2 + 
               \textstyle{1 \over 2} (1-k) H_-,             \nn \\
  & & O^{\dagger}O = H_-^3 - \textstyle{1 \over 2} (3+k) H_-^2 + 
               \textstyle{1 \over 2} (1+k) H_-,            \label{prod}
\eea
and
\beq
  [ O, \; O^{\dagger} ] = 3 H_-^2 -k H_-.                  \label{ten}
\eeq
RHS of (\ref{ten}) is not linear but quadratic. In order to treat it as 
linear algebra (Lie algebra), we must regard $H_-^2$ as a new element and 
further consider additional commutation relations. 
The commutators among $H_-^2$, $O$ and $O^{\dagger}$ yield new elements 
$O^{\dagger}H_-$ and $OH_-$. $ [O^{\dagger}H_-,\; O^{\dagger}] $ creates 
$ (O^{\dagger})^2 $, and in general $ (O^{\dagger})^{n+1} $ follows 
from $ [O^{\dagger}H_-, (O^{\dagger})^n ] $. An infinite number of 
commutation relations is thus brought about. The fundamental elements of our 
algebra are represented as 
$ \{ (O^{\dagger})^m H_-^n, \ O^m H_-^n, \ \ m, n = 0, 1, 2, \cdots \} $, 
to which elements such as $ O^n (O^{\dagger})^m $ are reduced because of 
(\ref{prod}). We refer to this algebra as $ \cal S $. Note that $k$ appears 
as a structure constant in (\ref{ten}).

  The purpose of this section is to discuss the relationship between  
algebras $ \cal S $ and \wi. It is convenient to first note 
that  algebra $ \cal S $ can be realized in terms of linear combinations 
of parts of $W_{1+\infty}$ generators; 
\bea
  (O^{\dagger})^m H_-^n & \rightarrow & 
  W(z^m (D+a_1 +m-1) (D + a_1 + m-2) \cdots (D+a_1) (D+a_0)^n)      \nn \\
  O^m H_-^n & \rightarrow & 
  W(z^{-m} (D+a_{-1}-m) (D+a_{-1}-m+1) \cdots (D+a_{-1}-1)    \label{mapw1} \\
  & & \hspace{0.7cm} \times(D-m+1) (D-m+2) \cdots D (D+a_0)^n),      \nn
\eea
where the possible values of $ a_{\pm 1} $ and $ a_0 $ are listed on Table~1. 
It is worth while noticing that these linear combinations are particular 
combinations which do not produce any central extensions whether or not the 
original \w1 generators participate in central extensions.

Let us show a brief sketch of how to determine $a_i$ and $\Psi=0$ for RHS 
of (\ref{mapw1}). Consider (\ref{seven}) and (\ref{ten}) in this realization; 
\beq
   H_- \rightarrow W(D+a_0), \qquad
   O^{\dagger}  \rightarrow W(z(D+a_1)), \qquad
   O \rightarrow W(z^{-1}(D+a_{-1}-1) D).    \label{mapsub}
\eeq
We see that (\ref{seven}) holds without any restriction on $a_i$ as usual.
\bea
 & & [H_-,\; O^{\dagger}] \rightarrow 
     [W(D+a_0),\; W(z(D+a_1))] = W(z(D+a_1)), \nn \\
 & & [H_-,\; O] \rightarrow 
     [W(D+a_0),\; W(z^{-1}(D+a_{-1}-1)D)] = -W(z^{-1}(D+a_{-1}-1)D). \nn
\eea
Comparing the image of (\ref{ten}) on both sides;
\bea
  [O,\; O^{\dagger}] &\rightarrow &
      [W(z^{-1}(D+a_{-1}-1) D),\; W(z(D+a_1)]              \nn \\
  &=& 3 W(D^2) + (2a_1 + 2a_{-1} - 1) W(D) + a_1 a_{-1}, \label{comoodag}
\eea
and
\bea
  3H_-^2 - kH_- &\rightarrow & 3 W((D+a_0)^2) - k W(D+a_0)              \nn \\
            &=& 3 W(D^2) + (6a_0-k) W(D) + 3a_0^2 - k a_0,  \label{rhs42}
\eea
we obtain the following two equations
\bea
 & & 2 a_1 + 2 a_{-1} = 6 a_0 + 1 -k,                   \nn \\
 & & a_1 a_{-1} = 3 a_0^2 - k a_0.                      \label{const}
\eea
Recalling that (\ref{prod}) is imposed on the definition of ${\cal S}$,  
these equations are not enough to conclude that the realization (\ref{mapw1}) 
is consistent with (\ref{prod}). We in fact obtain one more relation for 
$a_i$ considering
\beq
 [O,\; O^{\dagger} H_- ]  = [O,\; O^{\dagger}] H_- + O^{\dagger} O
  = (3H_-^2 -kH_-) H_- + \sigma (H_-),                           \label{oodagh}
\eeq
where $ \sigma (H_-) = O^{\dagger} O $ is given by RHS of (\ref{prod}). 
Comparing the coefficients of $ W(D^2) $ and $ W(D) $ on LHS 
of (\ref{oodagh})
\bea
  [O,\; O^{\dagger} H_-] & \rightarrow & 
  [W(z^{-1} (D+a_{-1}-1) D),\; W(z(D+a_1) (D+a_0))]   \nn \\
  &=& 4 W(D^3)  + 3(a_1 + a_{-1} + a_0 -1) W(D^2)     \label{new1} \\
  & & + (2 a_1 a_{-1} + (2a_0-1) (a_1 + a_{-1}) -a_0 + 1) W(D) 
     + a_0 a_1 a_{-1},                                      \nn 
\eea
with those on RHS
\bea
  & & (3H_-^2 -kH_-) H_- + \sigma (H_-)    \nn \\
  &\rightarrow & 4 W(D^3) + {1 \over 2} (24 a_0 - 3 -3k) W(D^2) 
  + ({1 \over 2} - 3a_0 + 12a_0^2 + {k \over 2} - 3 a_0 k) W(D) \nn \\
  & & + {1 \over 2}(a_0 - 3 a_0^2 + a_0 k -  3 a_0^2 k) + 4 a_0^3, \label{new2}
\eea
we obtain exactly the same relations as (\ref{const}). A comparison 
between constant terms gives another relation 
\beq
  2 a_0^3 - (k+3) a_0^2 + (k+1) a_0 = 0.        \label{const2}
\eeq
No further constraint is produced from 
$[O^m H_-^n,\; (O^{\dagger})^m H_-^l ]$, and we can determine 
$ a_i $ as solutions of (\ref{const}) and (\ref{const2}) as a result.

The following shows that the central extension (2-cocycle) 
of (\ref{mapw1}) always vanishes. 
The 2-cocycle that could appear only for 
$ [O^m H_-^n,\; (O^{\dagger})^m H_-^l ] $, $ (m \geq 1) $ is calculated;
\beq
  \Psi(z^{-m} r(D), z^m s(D)) = - \sum_{j=1}^m r(m-j) s(-j),   \label{van}
\eeq
with
\bea
  r(D) &=& (D+a_{-1}-m) (D+a_{-1}-m+1) \cdots (D+a_{-1}-1)    \nn \\
       &\times & (D-m+1) (D-m+2) \cdots D(D+a_0)^n,              \nn \\
  s(D) &=& (D+a_1+m-1)(D+a_1+m-2) \cdots (D+a_1)(D+a_0)^l,       \nn
\eea
and then
\bea
 \Psi & = & -\sum_{j=1}^m \;
(a_0+m-j)^n (a_0-j)^l (a_{-1}-j) (a_{-1}-j+1) \cdots (a_{-1}+m-j-1)   \nn \\
 & & \hspace{0.6cm} \times  (a_1+m-j-1) (a_1+m-j-2) 
 \cdots (a_1-j) (1-j) (2-j) \cdots (m-j)    \nn \\
 & = & 0.   \label{because}
\eea
This result is independent of the choice of $a_i$ because the last equality 
of (\ref{because}) is due to the factor $ (1-j) (2-j) \cdots (m-j) $.  
Obviously, this is consistent with the original fact that the generators of 
${\cal S}$ are combinations of $x$ and ${d \over dx}$. 
As  seen from (\ref{mapw1}), the mapping is one-to-one, however it is 
{\em not onto} because the particular elements 
$\{ W(z^{-m})\ |\  m \in {\rm \bf Z_{\geq 1}} \}$,  
which generate central extensions of \w1, are missing. This is also a 
reason that the realization of ${\cal S}$ is irrespective of central 
extensions of original \w1 generators.


Now, we give some remarks in the following. We point out there 
exists a one-to-one and onto relation between ${\cal S}$ 
and a subset of \wi.  
Eliminating also the generators 
associated to positive powers of $z$ 
$ \{ W(z^m) \ | \ m = \pm 1, \pm 2, \pm 3, \cdots \} $, for example, 
putting $ a_0 = a_1 = 0, a_{-1} = (1-k)/2 $, we find 
\bea
  (O^{\dagger})^m H_-^n  & \rightarrow &  
     W(z^m (D+m-1) (D+m-2) \cdots (D+1) D^{n+1}),  \nn \\
  O^m H_-^n  & \rightarrow &  
     W(z^{-m} (D+a_{-1}-m) (D+ a_{-1} -m + 1) \cdots (D+a_{-1}-1)
                                                  \label{mapwi} \\
   & & \hspace{0.6cm}
   \times (D-m+1) (D-m+2) \cdots (D-1) D^{n+1}).        \nn
\eea                                                                      

Secondly, similar to the Virasoro operators of the harmonic oscillator, 
$l_n=(b^{\dagger})^{n+1}b$ or $b^{\dagger}b^{n+1}$ for $n\geq 0$, 
we simply write down the Virasoro operators, which create/annihilate $n$ 
quanta for an excited state of $H_-$
\beq
  L_n = A^{\dagger} (b^{\dagger})^n A,
  \qquad   L^{\dagger}_n = A^{\dagger} b^n A.           \label{ncre}
\eeq
These satisfy the centerless Virasoro algebra of positive modes
\beq
  [ L_n,\; L_m ] = (m-n) L_{n+m},\qquad  
  [ L^{\dagger}_n,\; L^{\dagger}_m ] = (n-m) L^{\dagger}_{n+m}, 
\eeq
and 
\bea
   & & L_n \ket{G} = L^{\dagger}_n\ket{G}=0              \nn \\
   & & L_n \ket{\psi_m} = \left[
       (m + \textstyle{1 \over 2} (k+1)) (m+n+\textstyle{1 \over 2} (k+1)) 
       \displaystyle{(m+n)! \over m!} \right]^{1/2} \ket{\psi_{m+n}}  \\  
                                                           \label{act1}
  & & L^{\dagger}_n \ket{\psi_m} = \left\{
   \begin{array}{ll}
   \left[
       (m + \textstyle{1 \over 2} (k+1)) (m-n+\textstyle{1 \over 2} (k+1)) 
       \displaystyle{m! \over (m-n)!} \right]^{1/2} \ket{\psi_{m-n}} 
      & n \leq m \\
    0 & {\rm otherwise} \\
   \end{array}
   \right.     
\eea

%
\setcounter{equation}{0}
\section{Generalization to Parasupersymmetric Quantum Mechanics}
\indent

The formalism developed so far can be generalized to  
PSQM systems. Let us recall the definition of  
PSQM (of order 2, for example) based on ref.\cite{rs}. PSQM is 
essentially a pair of SQM Hamiltonians (except for vacuum structure)
\bea
  & & A_{\alpha} = {1 \over \sqrt 2} (ip + w_{\alpha}(x)), \qquad
     A_{\alpha}^{\dagger} = {1 \over \sqrt 2} (-ip + w_{\alpha}(x)), \qquad
     \alpha = 1, 2   \nn \\
  & & H_{SUSY}^{(1)} = \left(
     \begin{array}{cc}
     A_1 A_1^{\dagger}  & 0                 \\
     0                  & A_1^{\dagger} A_1  
     \end{array} \right)_,                               \label{psusy} \\
  & & H_{SUSY}^{(2)} = \left(
     \begin{array}{cc}
     A_2 A_2^{\dagger}  & 0                 \\
     0                  & A_2^{\dagger} A_2  
     \end{array} \right)_,                                \nn
\eea
equipped with the condition
\beq
   w_2^2 + w_2' = w_1^2 - w_1' + q,                     \label{shape}
\eeq
where $q$ is a constant. This is called the 
"shape-invariant condition" \cite{gen}. 
Because of condition (\ref{shape}), the PSQM of order 2 consists of three 
distinct Hamiltonians, {\em i.e.}
\bea
   & & H_1 = A_1 A_1^{\dagger},                 \nn \\
   & & H_2 = A_1^{\dagger} A_1 = A_2 A_2^{\dagger} - {q \over 2}, 
                                                \label{pham} \\
   & & H_3 = A_2^{\dagger} A_2.                 \nn
\eea
The Hamiltonians are isospectral except the states annihilated by 
$ A_{\alpha} $, since 
$ H_1 $ and $ H_2 $, $ H_2 $ and $ H_3 $ form SQM's respectively. 
The eigenstates of these Hamiltonians are transformed as follows
\[
\begin{array}{ccccc}
{\rm eigenstates}   & \stackrel{A_1^{\dagger}}{\longrightarrow} & 
{\rm eigenstates} & \stackrel{A_2^{\dagger}}{\longrightarrow} & 
{\rm eigenstates}  \\
{\rm of}\ H_1        
& \stackrel{\textstyle \longleftarrow}{\scriptstyle A_1} & {\rm of}\ H_2 &
\stackrel{\textstyle \longleftarrow}{\scriptstyle A_2} & {\rm of}\ H_3
\end{array}
\]

  To give the Hamiltonians equally spaced eigenvalues,  we require that 
 $ H_1 $ is the harmonic oscillator (similarly to sect.3)
\beq
   w_1^2 + w_1' = x^2 + k,                         \label{hampq}
\eeq
where $ k $ is a constant. The Hamiltonians $ H_1 $ and $ H_2 $ are  
identical to $ H_+ $ and $ H_- $ respectively. Hence we have another 
nonlinear differential equation (\ref{shape}) for each solution of 
the differential equation (\ref{hampq}). This determines $H_3$ as a new 
Hamiltonian with equally spaced eigenvalues. 
The previous SQM argument between $H_1$ and $H_2$ applies another SQM 
system of $H_2$ and $H_3$. Instead of $O$ and $O^{\dagger}$, we have 
\beq
  P = A_2^{\dagger} O A_2, \qquad
  P^{\dagger} = A_2^{\dagger} O^{\dagger} A_2,          \label{rlpq}
\eeq
and
\beq
 [H_3,\; P^{\dagger}] = P^{\dagger},\qquad [H_3,\; P] = -P, \label{capq}
\eeq
where $ O $ and $ O^{\dagger} $ are the raising and lowering operators 
for $ H_2 = H_- $ given in (\ref{six}). 
The counterparts to (\ref{prod}) and (\ref{ten}) are 
\bea
 PP^{\dagger} &=& H_3^5 - {1 \over 2} (3q+k-5) H_3^4 
+ {1 \over 4}(3q^2 + 2(k-6)q + 4(2-k)) H_3^3              \nn \\
 &-& {1 \over 8}(q^3+(k-9)q^2 -2(3k-7)q + 4(k-1))H_3^2 \label{pp1} \\
 &-& {1 \over 8}(q^3+(k-3)q^2 -2(k-1)q)H_3,              \nn \\
 P^{\dagger} P &=& H_3^5 - {1 \over 2}(3q+k+5) H_3^4 
+ {1 \over 4}(3q^2+2(k+6)q+4(k+2))H_3^3                   \nn \\
 &-& {1 \over 8}(q^3+(k+9)q^2+2(3k+7)q+4(k+1))H_3^2    \label{pp2} \\
 &+& {1 \over 8}(q^3 + (k+3)q^2+2(k+1)q)H_3,               \nn
\eea
and
\bea
[P, \; P^{\dagger}] &=& 5H_3^4 -2(3q+k) H_3^3 
+ ({9 \over 4}q^2 + {3 \over 2} kq + 1) H_2^2            \nn \\
&-& {q \over 4}(q^2+kq+2) H_3.                         \label{pp3}
\eea
Note that RHS of the commutator (\ref{pp3}) is in turn biquadratic compared to 
(\ref{ten}) and that two energy gap parameters appear in the structure 
constants. The algebra generated by $P$, $P^{\dagger}$ and $H_3$ 
is again infinite dimensional and its elements are 
$
\{P^n H_3^l, \ (P^{\dagger})^m H_3^l, \vert n,m,l \in{\bf \rm Z}_{\ge0}\}.
$ 
We refer to this algebra as \spq. 

 Similar to (\ref{mapwi}), the algebra \spq can be realized in terms of 
 \wi generators with central extensions. 
The realization is given by (one-to-one and onto)
\beq
  P^n H_3^m  \rightarrow  W(f(D)^n D^m), \qquad
  (P^{\dagger})^n H_3^m \rightarrow W(g(D)^n D^m),     \label{spsmap}
\eeq
where $ f(D) $ and $ g(D) $ are defined by
\[
   f(D) = z^{-1} (D^2+\alpha D + \beta) D, \qquad
   g(D) = z(D+ \lambda) D,
\]
and the 2-cocycle vanishes for $[P^n H_3^l, \; (P^{\dagger})^n H_3^m]$.
Possible values of $ \alpha, \beta $ and $ \lambda $ (listed on Table~2) 
are determined through comparing the image on both sides of (\ref{pp3}) 
\bea
 & & [P,\; P^{\dagger}] \ \rightarrow \ [ f(D), \; g(D) ] \nn \\
 & & = 5 W(D^4) + (4 \alpha + 4 \lambda + 2) W(D^3) 
  + (3 \alpha \lambda + 3 \lambda + 3 \beta + 1) W(D^2) \label{pp5} \\
 & & + (-\beta + \lambda + \alpha \lambda + 2 \beta \lambda ) W(D), \nn
\eea
\bea
  & & \chi(H_3) \rightarrow 5 W(D^4) - 2(3q+k) W(D^3) 
     + ({9 \over 4}q^2 + {3 \over 2} kq + 1) W(D^2)  \label{new3} \\
  & & - {q \over 4} (q^2 + kq + 2) W(D),   \nn
\eea
where $ \chi(H_3) $ is the RHS of (\ref{pp3}).  
We thus obtain the equations
\bea
 & & 2 \alpha + 2 \lambda + 1 = -3q -k, \nn \\
 & & 4 \beta + 4 \lambda + 4 \alpha \lambda = 3 q^2 + 2 qk,  \label{eqs1} \\
 & & -4 \beta + 4 \lambda + 4 \alpha \lambda + 8 \beta \lambda 
     = -q^3 - kq^2 - 2q. \nn
\eea
These equations (\ref{eqs1}) are consistent also with the image of the 
commutation relation 
\[
  [P,\; P^{\dagger} H_3] = [P,\; P^{\dagger}]H_3 + P^{\dagger}P 
  = \chi(H_3) H_3 + \rho(H_3), 
\]
where $ \rho(H_3) = P^{\dagger} P $ is given by RHS of (\ref{pp2}). 

\section{Conclusion}
\indent

In the present paper, we discussed  spectrum generating algebras 
of the SQM and PSQM systems with equally spaced eigenvalues. (P)SQM 
Hamiltonians with equally spaced eigenvalues can be obtained for each solution 
to the nonlinear differential equations (\ref{five}) and (\ref{shape}). The 
pair of Hamiltonians is solvable; namely, we can construct all of their 
eigenstates using the raising/lowering operators in each system. 
One satisfies a 
harmonic oscillator algebra and the other can be realized by \wi with 
an arbitrary value of $c$.  
The commutation relations (\ref{seven}) are common with harmonic oscillator 
algebra, while eq.(\ref{ten}) is different. One can discuss 
different algebraic aspect from ours; for example, 
Fern\'andez et al modify the definition of $ O $ and $ O^{\dagger} $ 
and obtained a boson-like commutation relation \cite{fnr}. They also 
constructed a coherent state for their annihilation operator. 


Our formalism is dependent on the factorization of Hamiltonians and on the 
choice of creation/annihilation operators for $H_+$, which we have assumed 
to be a harmonic oscillator. 
There exist other types of Hamiltonian with equally 
spaced eigenvalues \cite{ng}. It is still an open question whether our 
formalism is applicable to such systems. As to PSQM, one might get 
more interesting results as a continuation of this work along the 
line of \cite{para}. 

\section*{Acknowledgements}

  This work was started from discussions by one of us (N. A.) with 
Professor H. Ui (Hiroshima University). We would like to express our 
thanks to him for enlightening discussions and suggestions.

%
\setcounter{equation}{0}
\section*{Appendix: examples of $w(x)$}
\indent

  Roy and Roychoudhury \cite{rr} found an infinite sequence of solutions of 
(\ref{five}) when $k=1$. That is given by
\bea
   & & w_{n+1}(x) = w_n(x) + u_n(x), \qquad 
       w_0 = x                       \qquad
       n = 0, 1, 2, \cdots                   \nn \\
   & & u_n = { \displaystyle{\exp (-2 \int^x w_n(t) dt) \over
       c_n + \int^x \exp (-2\int^{x'} w_n(t) dt) dx' }}, \label{5teen}
\eea
where $ c_n $ is the constant determined so as to make the 
ground state normalizable. The Hamiltonian corresponding to $ w_n(x) $ 
is given by
\beq
  H_-^{(n)} = {1 \over 2} (p^2+x^2-1) - \sum_{m=0}^{n-1} {du_m(x) \over dx}.
                                                        \label{6teen}
\eeq
Let us look at the case of $ n = 1 $ in detail. The Hamiltonian of 
this case
\beq
  H_-^{(1)} = {1 \over 2} (p^2+x^2-1) - {du_0(x) \over dx},
  \qquad
  u_0(x) = {\exp(-x^2) \over c_0 + \int_{-\infty}^x \exp(-t^2) dt }, 
                                                        \label{7teen}
\eeq
is already discussed by several authors \cite{bm,am,ng}. The 
ground state of this Hamiltonian is given by 
\beq
    \ket G = \left( {c_0(c_0 + \sqrt{\pi}) \over \sqrt{\pi}} \right)^{1/2} 
        { \exp(-x^2/2) \over c_0 + \int_{-\infty}^x \exp(-t^2/2) dt}.
                                                       \label{8teen}
\eeq
The normalizability of the ground state requires that the constant 
$ c_0 $ lies in the range 
$ c_0 > 0 $ or $ c_0 < -\sqrt{\pi}. $ The shape of potential depends on 
the value of $ c_0 $, and hence we have obtained uncountable infinite 
number of solutions of the nonlinear differential equation (\ref{five}). 
The potentials for some values of $ c_0 $ are depicted in Figure 2.

%

\newpage

\section*{Figure Captions}

{\bf Figure ~1 : }\ Energy spectra of $ H_{\pm} $ and action of operators

\noindent
{\bf Figure~2 : }\ Plots of the potential 
$ V(x) = \displaystyle{x^2 \over 2} - {du_0(x) \over dx} $ 
for various values of $ c_0. $ The thin and thick solid lines correspond to 
$c_0 = -2,\ 0.1$, while the dotted and dashed lines to $c_0 = 2,\ 10$
respectively.

\newpage

\setlength{\unitlength}{1cm}

\begin{figure}[h]
\begin{picture}(12,12)(-4,1)
\put(0,0){\line(1,0){9}}  
\put(0,0){\vector(0,1){10}} 
\linethickness{1mm}
\thicklines
\put(2,0){\line(1,0){1.5}}  
\multiput(2,2)(0,1.5){5}{\line(1,0){1.5}} 
\multiput(6,2)(0,1.5){5}{\line(1,0){1.5}} 
\thinlines
\multiput(2.4,2.25)(0,1.5){4}{\vector(0,1){1}}   
\multiput(6.4,2.25)(0,1.5){4}{\vector(0,1){1}}   
\multiput(3.1,7.75)(0,-1.5){4}{\vector(0,-1){1}} 
\multiput(7.1,7.75)(0,-1.5){4}{\vector(0,-1){1}} 
\put(4,4.5){\vector(1,0){1.5}}                   
\put(5.5,5.5){\vector(-1,0){1.5}}                
%
%
\put(-0.5,10){$E$}
\put(-0.5,-0.2){$ 0 $}
\put(1,0.2){$ \ket G $}
\newcounter{row}
\setcounter{row}{0}
\multiput(0.9,1.85)(0,1.5){5}{
$ \ket{\psi_{\arabic{row}} \addtocounter{row}{1} } $
}
\setcounter{row}{0}
\multiput(7.7,1.85)(0,1.5){5}{
$ \ket{\arabic{row}} $ \addtocounter{row}{1} 
}
\put(-1,1.8){$ E_0 $}
\setcounter{row}{1}
\multiput(-1.5,3.3)(0,1.5){4}{
$ E_0 + \arabic{row} $ \addtocounter{row}{1} 
}
\put(2.2,8.2){$ O^{\dagger}$}
\put(2.9,8.2){$ O $}
\put(6.3,8.2){$ b^{\dagger} $}
\put(7.0,8.2){$ b $}
\put(4.7,4){$ A $}
\put(4.7,5.8){$ A^{\dagger} $}
\put(2.7,-0.5){$ H_- $}
\put(6.7,-0.5){$ H_+ $}
\end{picture}
\vspace*{3cm}
\caption{Energy spectra of $ H_{\pm} $ and action of operators}
\end{figure}
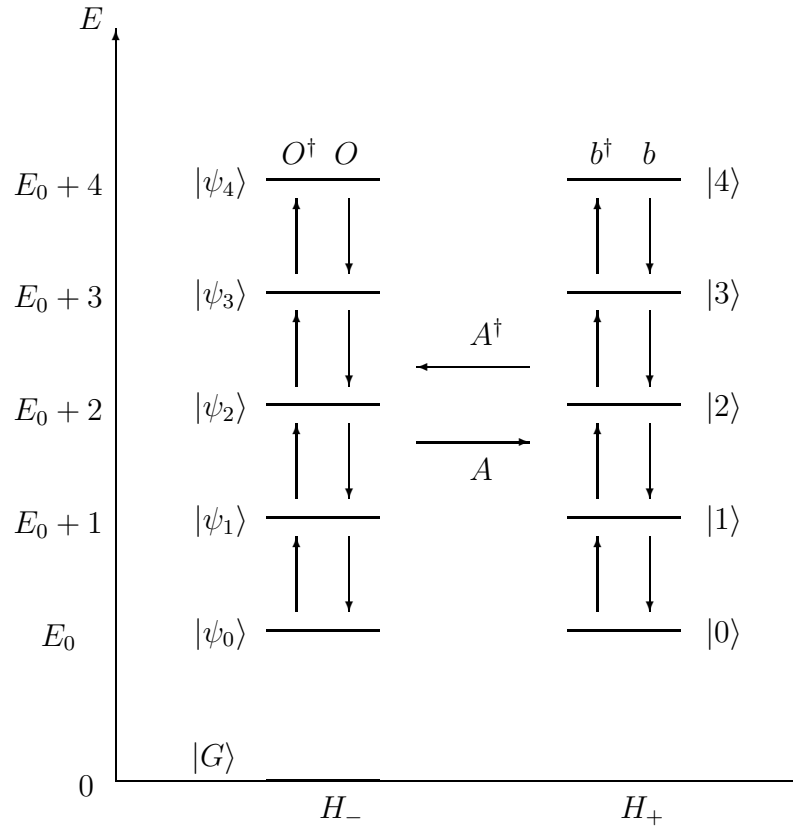

\newpage 

\begin{center}
\begin{tabular}{|ccc|} \hline
$ a_0 $ & $ a_1 $ & $ a_{-1} $ \\ \hline 
\hline
0 &
  \begin{tabular}{c} 
  0 \\
  $ {1 \over 2} (1-k) $
  \end{tabular}
&
  \begin{tabular}{c}
  $ {1 \over 2} (1-k) $ \\
  0
  \end{tabular}
\\ \hline
1 &
  \begin{tabular}{c} 
  2 \\
  $ {1 \over 2} (3-k) $
  \end{tabular}
&
  \begin{tabular}{c}
  $ {1 \over 2} (3-k) $ \\
  2
  \end{tabular}
\\ \hline
$ {1 \over 2} (k+1) $ &
  \begin{tabular}{c} 
  $ {1 \over 2} (k+1) $ \\
  $ {1 \over 2} (k+3) $
  \end{tabular}
&
  \begin{tabular}{c}
  $ {1 \over 2} (k+3) $ \\
  $ {1 \over 2} (k+1) $
  \end{tabular}
\\ \hline
\end{tabular}

\vspace{2cm}
{\bf Table~1 : } \ allowed values of constants $ a_0 $ and $ a_{\pm} $
\end{center}

\newpage

\begin{center}
\begin{tabular}{|ccc|} \hline
$ \alpha $ & $ \beta $ & $ \lambda $ \\ \hline 
\hline
$ \displaystyle{-{1 \over 2}(2q+k+1)} $ 
& $ \displaystyle{{1 \over 4}(q^2+(k+1)q)} $ 
& $ \displaystyle{-{q \over 2}} $ \\
$ \displaystyle{-{1 \over 2}(2q+k+3) } $ 
& $ \displaystyle{{1 \over 4}(q^2+(k+3)q+2(k+1))} $
& $ \displaystyle{{1 \over 2}(-q+2)} $ \\
$-q-1$  
& $ \displaystyle{{1 \over 4}(q^2+2q)} $ 
& $ \displaystyle{{1 \over 2}(-q-k+1)} $
\\ \hline
\end{tabular}

\vspace{2cm}
{\bf Table~2 : } \ allowed values of constants $ \alpha, \beta $ 
and $ \lambda $
\end{center}
\end{document}